\documentclass{ws-procs9x6}            
\begin{document}
\title{Multiple charm and hidden charm mesons with strangeness}

\author{X. L. Ren}

\address{Ruhr-Universit\"{a}t Bochum, Fakult\"{a}t f\"{u}r Physik und Astronomie, Institut f\"{u}r Theoretische Physik II, D-44780 Bochum, Germany.}

\author{Brenda B. Malabarba}
\address{Universidade de Sao Paulo, Instituto de Fisica, C.P. 05389-970, Sao Paulo, Brazil.}

\author{L. S. Geng}
\address{School of Physics \& Beijing Key Laboratory of Advanced Nuclear Materials and Physics, Beihang University, Beijing 100191, China.}
\address{Beijing Advanced Innovation Center for Big Date-based Precision Medicine, Beihang University, Beijing100191, China.}
\author{K. P. Khemchandani}
\address{Universidade Federal de Sao Paulo, C.P. 01302-907, Sao Paulo, Brazil.}

\author{A. Mart\'inez Torres$^*$}
\address{Universidade de Sao Paulo, Instituto de Fisica, C.P. 05389-970, Sao Paulo, Brazil.\\
$^*$E-mail: amartine@if.usp.br}

\begin{abstract}
In this talk we summarize our latest results on the study of three-body systems with explicit/hidden charm and with strangeness. In particular, we focus on the $K D D $ and
$K D\bar D^*$ systems, where a charm $+2$, isospin $1/2$ and strangeness $+1$ state is found with a mass around 4140 MeV in the former and a $K^*$ state, with hidden charm, at a mass around 4307 MeV is obtained in the latter. Both states are predictions of our model and the experimental confirmation of them would be important to understand the properties of the strong interaction in the presence of heavy quarks.
\end{abstract}

\bodymatter

\section{Introduction}
In the present understanding of Quantum Chromodynamics, nature permits the existence of mesons and baryons with several units of charm quantum number. Some examples are the doubly charmed mesons and baryons $T_{cc}$, $\Xi^+_{cc}$, $\Omega^+_{cc}$ and triply charmed baryons arising from the dynamics involved in systems like $\Xi_{cc} D$, $\Xi_{cc} D^*$, $\Xi_{cc}\Lambda_c$, $\Xi_{cc}\Sigma_c$~\cite{Mattson:2002vu,Zouzou:1986qh,Chen:2017jjn,Chen:2018pzd}. In the last years, the interest of investigating systems with charm and bottom quantum numbers has been extended to three-body systems, like $BD\bar D$, $B DD$, $BBB^*$, $DDK$~\cite{Dias:2017miz,Ma:2017ery,SanchezSanchez:2017xtl}. However, an experimental investigation of these states is necessary to corroborate the available information on their properties from theoretical works. And such a research proposal is presently missing in the projects of the different experimental facilities. The search for this kind of states would definitely shed light on the working of the strong interaction in the presence of heavy quarks and can open new research lines, like the search of excited states of kaons and/or $K^*$'s at energies around 4-5 GeV with hidden charm and/or bottom quantum numbers. 

In this talk we show our results for two kind of three-body systems with strangeness, the $KDD$ system, where we find a state with mass around 4140 MeV, in agreement with the findings of Ref.~\cite{SanchezSanchez:2017xtl}, based on a different formalism, and the $KD\bar D^*$ system, where we find the existence of a $K^*$ vector meson with hidden charm and mass around 4307 MeV, as also found in Ref.~\cite{Ma:2017ery}.

\section{Formalism and Results}
To study the formation of states in three-body systems like $KDD$ and $KD\bar D^*$ we solve the Faddeev equations using as kernel two-body $t$-matrices obtained from the resolution of the Bethe-Salpeter equation in its on-shell factorization form~\cite{Oller:1997ti,Oset:1997it}. The advantage of the latter is that the Bethe-Salpeter equation, which is given by
\begin{align}
t=t+vgt,\label{BS}
\end{align}
where $v$ is the kernel and $g$ is a loop function of two hadrons, becomes an algebraic equation, which is solved by considering two-body coupled channels. Using these $t$-matrices, we solve the Faddeev equations~\cite{Faddeev:1960su}
\begin{align}
T^1&=t^1+t^1 G[T^2+T^3],\nonumber\\
T^2&=t^2+t^2 G[T^1+T^2],\nonumber\\
T^3&=t^3+t^3 G[T^1+T^2],\label{Fa}\
\end{align}
within the approach developed in Refs.~\cite{MartinezTorres:2007sr,Khemchandani:2008rk,MartinezTorres:2010zv} for the case of the $DDK$ system, and that developed in Refs.~\cite{Xie:2010ig,Xie:2011uw,MartinezTorres:2010ax} for the $KD\bar D^*$ system.

To determine the kernel $v$ in Eq.~(\ref{BS}) we have used effective Lagrangians based on the chiral and heavy quark symmetries of the strong interaction~\cite{Gasser:1982ap,Gasser:1984gg,Georgi:1990um,Isgur:1991wq,Wise:1992hn,Burdman:1992gh}. It is now widely accepted that the non-perturbative coupled channel dynamics generates the state $D^*_{s0}(2317)$ in the $KD$ system~\cite{Kolomeitsev:2003ac,Guo:2006fu,Gamermann:2006nm,Torres:2014vna,Altenbuchinger:2013vwa}, while in the $D\bar D^*$ system we have the formation of the isospin 0 state $X(3872)$ and of the isospin 1 state $Z_c(3900)$~\cite{Close:2003sg,Wong:2003xk,Swanson:2003tb,Gamermann:2007fi,Aceti:2014uea}.

With these ingredients, the solution of Eq.~(\ref{Fa}) for the $DDK$, $DD\pi$ and $DD_s\eta$ coupled channel system reveals the formation of a bound state with isospin 1/2, charm +2, strangeness +1 and mass around 4140 MeV~\cite{MartinezTorres:2018zbl}. The state is found to arise when one of the $DK$ subsystems, together with its coupled channels, generate the $D^*_{s0}(2317)$. This result is similar to the one found in Ref.~\cite{SanchezSanchez:2017xtl} in which the system $D-D^*_{s0}(2317)$ was studied by means of an effective two-body potential. We have also estimated the size of such an exotic state to know whether it is compact or not. To do this, we have treated the state found as a $D-D^*_{s0}(2317)$ state and estimate the root mean square (RMS) distance among the constituent hadrons~\cite{MartinezTorres:2018zbl} by constructing the wave function of the state, for which we follow Ref.~\cite{Gamermann:2009uq}. We find the RMS$\sim 1.0-1.4$ fm, which is larger than that of $D^*_{s0}(2317)$~\cite{Torres:2014vna,MartinezTorres:2018zbl}. This result has been also confirmed by the authors of Ref.~\cite{Wu:2019vsy}.  The state found in Refs.~\cite{SanchezSanchez:2017xtl,MartinezTorres:2018zbl} is a three-body bound state, thus, it has a zero width. However, a state with these properties can decay via triangular loops to two-body channels like $D_s D$, $D_s D^*$ and $D D^*_s$ producing a small width. We have also studied such decay mechanisms in Ref.~\cite{Huang:2019qmw}, finding a total width for the state to be around 2-3 MeV.

In case of the $KD\bar D^*$ system, the resolution of Eq.~(\ref{Fa}) shows the generation of a $K^*$ meson with mass around 4307 MeV when the $KD$ system in isospin 0 forms the $D^*_{s0}(2317)$ and the interaction in the $D\bar D^*$ system give rise to $X(3872)$ in isospin 0 and $Z_c(3900)$ in isospin 1~\cite{Ren:2018pcd}. This three-body state, as in case of the one found in the $DDK$ system, appears below the $KD\bar D^*$ threshold, thus, it is a bound state and has zero width. However, the $Z_c(3900)$, considering it to be a state generated from the $D\bar D^*$ and $J/\psi\pi$ coupled channels, has a width of around 30 MeV from its decay to $J/\psi\pi$~\cite{Tanabashi:2018oca}. When considering this width into the calculation, the three-body state found in the $KD\bar D^*$ system gains a width of around 18 MeV. This three-body state can also decay to two-body channels, like $J/\psi K^*(892)$, $\bar D D_s$, $\bar D D^*_s$ and $\bar D^* D^*_s$ via triangular loops. In Ref.~\cite{Ren:2019umd} we have obtained these decay widths, finding the values $\sim 7$ MeV for $J/\psi K^*(892)$, $\sim 0.5$ for $\bar D D^*_s$, $\bar D^* D^*_s$ and $\sim 1$ MeV for $\bar D D_s$. A three-body state like this, with a dominant $KZ_c(3900)$ component, can also easily decay to a state formed by $J/\psi\pi K$, thus, the reconstruction on the $J/\psi\pi K$ invariant mass in reactions involving these particles in their final states can be a way of finding this $K^*(4307)$ experimentally. 

\section{Conclussions}
We have presented in this talk our latest results for the $KDD$ and $KD\bar D^*$ systems and the formation of exotic states in them whose properties can be understood as molecular kind of states. In particular, in the $KDD$ system we find a state around 4140 MeV and in the $KD\bar D^*$ system we obtain a $K^*$ vector meson with mass around 4307 MeV.

\section*{Acknowledgments}
This work was partly supported the National Natural Science Foundation of China (NSFC) under Grants Nos. 11735003 and 11522539, by DFG and NSFC through funds provided to the Sino-German CRC 110 ``Symmetries and the Emergence of Structure in QCD'' (Grant No. TRR110), and Conselho Nacional de Desenvolvimento Cient\'ifico e Tecnol\'ogico (CNPq) under Grant Nos. 310759/2016-1 and 311524/2016-8. A. M. T gratefully acknowledges the support from the program ``Mobilidade Santander - Docentes'' (Edital PRPG-11/2019).

\end{document}